\documentclass[preprint,authoryear,12pt]{elsarticle}

\usepackage{amssymb}
\usepackage{amsmath}

\journal{Earth and Planetary Science Letters}

\usepackage[usenames]{color}

\newcommand{\new}[1]{#1}

\newcommand{\f}[3]{$_{#2\,}#1^{\,#3}$}
\newcommand{\fpm}[3]{\ensuremath{#1_{\,#2}^{\,#3}}}
\newcommand{\eanu}{\ensuremath{\bar\nu_e}}
\newcommand{\cmms}{\ensuremath{\text{cm}^{-2}\,\mu\text{s}^{-1}}}
\newcommand{\Unum}{\ensuremath{^{238}\text{U}}}
\newcommand{\Ufive}{\ensuremath{^{235}\text{U}}}
\newcommand{\Tnum}{\ensuremath{^{232}\text{Th}}}
\newcommand{\Knum}{\ensuremath{^{40}\text{K}}}
\newcommand{\rr}{\mathbf r}
\newcommand{\RG}{R\&G}
\newcommand{\WH}{W\&H}
\newcommand{\SaS}{S\&S}
\newcommand{\AMcD}{A\&McD}
\newcommand{\WK}{W\&K}

\begin{document}

\begin{frontmatter}



\title{Geophysical and geochemical constraints on geoneutrino fluxes from Earth's mantle}

\author[umd]{Ond\v rej \v Sr\'amek\corref{cor1}}
\ead{sramek@umd.edu}
\author[umd]{William F. McDonough}
\author[caltech]{Edwin S. Kite}
\author[umd]{Vedran Leki\'c}
\author[hpu,uhi]{Stephen T. Dye}
\author[ucb]{Shijie Zhong}

\cortext[cor1]{Corresponding author}

\address[umd]{Department of Geology, University of Maryland, College Park, MD 20742, USA}
\address[caltech]{Division of Geological and Planetary Sciences, California Institute of Technology, Pasadena, CA 91125, USA}
\address[hpu]{Department of Natural Sciences, Hawaii Pacific University, Kaneohe, HI 96744 USA}
\address[uhi]{Department of Physics and Astronomy, University of Hawaii, Honolulu, HI 96822 USA}
\address[ucb]{Department of Physics, University of Colorado at Boulder, Boulder, CO 80309, USA}

\begin{abstract}
Knowledge of the amount and distribution of radiogenic heating in the mantle is crucial for understanding the dynamics of the Earth, including its thermal evolution, the style and planform of mantle convection, and the energetics of the core. Although the flux of heat from the surface of the planet is robustly estimated, the contributions of radiogenic heating and secular cooling remain poorly defined. Constraining the amount of heat-producing elements in the Earth will provide clues to understanding nebula condensation and planetary formation processes in early Solar System. Mantle radioactivity supplies  power for mantle convection and plate tectonics, but estimates of mantle radiogenic heat production vary by a factor of \new{more than 20}. Recent experimental results demonstrate the potential for direct assessment of mantle radioactivity through observations of geoneutrinos, which are emitted by naturally occurring radionuclides. Predictions of the geoneutrino signal from the mantle exist for several established estimates of mantle composition. Here we present novel analyses, illustrating surface variations of the mantle geoneutrino signal for models of the deep mantle structure, including those based on seismic tomography. These variations have measurable differences for some models, allowing new and meaningful constraints on the dynamics of the planet. An ocean based geoneutrino detector deployed at several strategic locations will be able to discriminate between competing compositional models of the bulk silicate Earth.
\end{abstract}


\end{frontmatter}


\section{Introduction}
\label{sec:intro}

The present-day Earth surface heat flux is $47\pm1$(stat.)\,TW based on latest analysis of \citet{davies:2010}, in agreement with recent estimate of $46\pm3$\,TW by \citet{jaupart:2007tg}. The two \new{main} contributors to the surface heat loss are secular cooling of the Earth, and heat generated by decay of long-lived radioactive isotopes of uranium, thorium, and potassium. The relative magnitude of these two components remain poorly constrained.
Estimates of the present-day heat-producing element (HPE) abundances in the bulk silicate Earth (BSE, defined as the entire Earth less its metallic core) vary by a factor of about three between different models \citep{oneill:2008,javoy:2010,arevalo:2009,TS:2002}. Compositional estimates of depleted mantle (DM), which is the source of mid-ocean ridge basalt (MORB), vary by a similar factor \citep{workman:2005,salters:2004,arevalo:2010}.
A distinct chemical reservoir is usually invoked to account for the apparent deficit of some elements and isotopes in the BSE chemical inventory \citep{hofmann:1997nat}. Enriched in HPEs and some other elements (e.g., helium, argon), and possibly ``hidden'' \citep[i.e., untapped by surface volcanism;][]{boyet:2006}, this reservoir is usually assumed to be located in the lowermost mantle. 
Because no methods exist for directly accessing and analyzing samples of Earth's deep mantle, compositional estimates rely on chemical analyses of available rock samples (coming from a relatively shallow mantle at best), interpretations of  indirect evidence from geophysical data (e.g., seismology), and a number of simplifying assumptions (e.g., relating Earth's composition to the Solar System or meteorite chemistry). Consequently, mass balances for different chemical elements often yield inconsistent estimates of the size and enrichment of the deep reservoir \citep{hofmann:1997nat}.

Recent advances in experimental neutrino physics provide a breakthrough in deep-Earth research. Geoneutrino detections by KamLAND \citep{araki:2005nat,gando:2011bis} and Borexino \citep{bellini:2010}, using land-based instruments, are consistent with flux predictions. These analyses assume a planetary Th/U ratio and absence of U and Th in the core.
Up to now, predictions of geoneutrino fluxes coming from the mantle consider spherically symmetric HPE distributions, including uniform mantle and layers of varying depth and thickness \citep{araki:2005nat,bellini:2010,gando:2011bis,mantovani:2004,enomoto:2007,fiorentini:2007,dye:2010}. 
However, global seismic tomography reveals two large, low shear velocity provinces (LLSVPs, also referred to as superplumes or thermochemical piles) at the base of the mantle beneath Africa and the Pacific. Sharp velocity gradients bound the LLSVPs \citep{wen:2001a}, suggesting a compositional difference from ambient lower mantle. This conclusion is supported by the observation that shear and sound wavespeeds are anti-correlated in the lowermost mantle \citep{su:1997}. 
Moreover, existing mantle geoneutrino predictions are usually based on a single compositional model, even though several estimates for both BSE and DM composition exist. 

Our new predictions of geoneutrino signal from the Earth's mantle recognize the latest geophysical constraints and consider several established compositional estimates for the Earth's reservoirs. In section~\ref{sec:fluxcalc} we introduce the calculation of geoneutrino flux. Estimates of HPE abundances in BSE and the crust are discussed in section~\ref{sec:hpe}. Section~\ref{sec:flux} presents predictions of geoneutrino emission from the mantle with various assumptions about HPE distribution, including a premise that seismically imaged deep-mantle structures may reflect a compositional difference. Section~\ref{sec:detect} focuses on detectability of predicted mantle flux lateral variations, followed by general discussion in section~\ref{sec:discussion}.

\section{Geoneutrino flux calculation}
\label{sec:fluxcalc}

Beta-decays in decay chains of radionuclides $\Unum$, $\Ufive$, $\Tnum$ and $\beta$-decay of $\Knum$ produce electron antineutrinos. The antineutrino flux $\Phi_X(\rr)$ at position $\rr$ from a radionuclide $X$ at positions $\rr'$ distributed in a spatial domain $\Omega$ is calculated from
\begin{equation} \label{flux0}
  \Phi_X(\rr) = \frac{n_X \lambda_X \langle P \rangle}{4\pi} \int\limits_\Omega \frac{a_X(\rr') \rho(\rr')}{|\rr-\rr'|^2} \mathrm{d}\rr',
\end{equation}
where $n_X$ is the number of antineutrinos per decay chain, $\lambda_X$ is the decay constant (1/lifetime), $a_X$ is the abundance of radioactive isotope (number of atoms of radioactive isotope per unit mass of rock), and $\rho$ is rock density \citep{mantovani:2004}. The average survival probability \new{$\langle P \rangle=\fpm{0.544}{-0.013}{+0.017}$ \citep{dye:2012rg}} assumes a signal source region size much larger than the neutrino oscillation length \citep[60--110\,km depending on antineutrino energy; see][for more extensive discussion]{dye:2012rg}.
The isotopic abundance $a_X$ is calculated from
\begin{equation} \label{abundance}
  a_X = \frac{A_XX_X}{M_X},
\end{equation}
where $A_X$ is the elemental abundance (mass of element per unit mass of rock), $X_X$ is the isotopic ratio (atoms of radionuclide per atoms of element), $M_X$ is atomic  mass. Radiogenic heating rate $H_X$ (power per unit mass of rock) by radionuclide $X$ is calculated from
\begin{equation} \label{heating}
  H_X = a_X \lambda_X Q^h_X,
\end{equation}
\new{where $Q^h_X$ is the energy, per decay of one atom of the parent radionuclide, available for radiogenic heating. It is the total decay energy less the fraction carried away by antineutrinos from $\beta$-decays. In the case of a decay chain, $Q^h_X$ sums the contributions from each $\alpha$- and $\beta$-decay in a decay chain \citep{dye:2012rg}.}
Values of atomic parameters in equations (\ref{flux0}--\ref{heating}) are listed in Table~\ref{tab:atomic}. Input from geochemistry and geophysics is required for the elemental abundances $A_X$ and rock density $\rho$. For a spherical shell source region with uniform rock density and uniform radionuclide abundance, the flux \eqref{flux0} can be evaluated analytically \citep{krauss:1984,fiorentini:2007}.

Current experimental methods for geoneutrino detection, which employ the neutron inverse $\beta$-decay reaction, are only able to detect the highest energy geoneutrinos from $\Unum$ and $\Tnum$ decay chains. The conversion factor between the signal (geoneutrino flux) and a measurement (number of detected events) is a function of the detector size (number of free target protons), experiment duration (live-time) and detection efficiency. A convenient ``terrestrial neutrino unit'' (TNU) was devised as 1 event detected over 1 year exposure of $10^{32}$ target protons at 100\% detection efficiency \citep{mantovani:2004}. One TNU corresponds to a flux of $7.67\times10^4$\,cm$^{-2}$\,s$^{-1}$ from $\Unum$ or $2.48\times10^5$\,cm$^{-2}$\,s$^{-1}$ from $\Tnum$ \citep{enomoto:2007}. The conversion for a combined signal from $\Unum$ and $\Tnum$ depends on the Th/U abundance ratio of the source; for $\text{Th/U}\approx4$ about 80\% of the measured events comes from $\Unum$ and the remaining 20\% from $\Tnum$ \citep[see, e.g.,][for detailed description]{dye:2012rg}.

\section{HPE abundances in BSE and the crust}
\label{sec:hpe}

Three classes BSE compositional estimates -- termed here ``cosmochemical'', ``geochemical'', and ``geodynamical'' -- give different abundances of HPEs. 
The cosmochemical approach bases Earth's composition on enstatite chondrites, which show the closest isotopic similarity with mantle rocks and have sufficiently high iron content to explain the metallic core \citep{javoy:2010}. Cosmochemical estimates suggest relatively low HPE abundances.
\new{Following \citet{javoy:2010}, we use bulk Earth uranium and thorium abundances of CI chondrites from \citet{wasson:1988}, $A_U=8.2\,(\pm20\%)$\,ppb and $A_{Th}=29\,(\pm10\%)$\,ppb \citep[consistent with EH Earth model of][]{javoy:1999}. We then multiply these values by the enrichment factor for refractory lithophile elements of 1.479 that accounts for the differentiation of an early Earth into core and mantle \citep{javoy:2010}, and get U and Th abundances in BSE of $12\pm2$\,ppb and $43\pm4$\,ppb. We consider a K/U ratio of 12000, leading to an abundance of the moderately volatile potassium of $A_K=146\pm19$\,ppm in BSE for the cosmochemical estimate.}

There are other low Earth models that have similarly low abundance of the heat producing elements. \citet{oneill:2008} recently proposed a model whereby the early Earth was developing a crust, enriched in highly incompatible elements (e.g., U, Th, and K) that experienced collisional erosion, which resulted in marked depletions of these elements from the bulk silicate Earth.  Consequently, the \citeauthor{oneill:2008} model has a bulk silicate Earth that contains as little as 10\,ppb\,U, 40\,ppb\,Th and 140\,ppm\,K, which is, in terms of absolute concentration, comparable to the \citeauthor{javoy:2010} model.

Geochemical estimates adopt chondritic compositions for the relative abundances of refractory lithophile elements with absolute abundances constrained by terrestrial samples \citep{mcdonough:1995}, and have moderate abundances of HPEs.
We use a geochemical estimate of \citet{arevalo:2009}, which is a modified version of \citeauthor{mcdonough:1995}'s (\citeyear{mcdonough:1995}) model. The uncertainties are included, and within the errors the proposed values are consistent with other geochemical estimates \citep{hart:1986,allegre:1995epsl,palme:2003tgc}.

Geodynamical estimates are based on the energetics of mantle convection and the observed surface heat loss \citep{TS:2002}. Classical parameterized thermal evolution models require a significant fraction ($\gtrsim60\%$) of the present-day mantle heat output to be contributed by radiogenic heating in order to prevent extremely high temperatures in Earth's early history, which is ruled out by geological observations. This is commonly expressed in terms of the mantle Urey ratio, defined as mantle radiogenic heat production over total heat output from the mantle. The mantle Urey ratio characterizes the energy available for mantle convection and plate tectonics, which is mostly accretional energy from Earth formation for $\text{Ur}<0.5$, and mostly ongoing radioactivity for $\text{Ur}>0.5$. 
Our geodynamical HPE abundance estimate is based on values of \citet{TS:2002}, scaled to result in mantle Urey ratio of 0.6--0.8.
Table~\ref{tab:abund} lists the U, Th, and K abundances and the Th/U and K/U mass ratios for the three BSE compositional estimates. 
\new{It is assumed that the uncertainties in U, Th, and K abundances are fully correlated.}
The rates of radiogenic heat production are $11\pm2$\,TW, $20\pm4$\,TW and $33\pm3$\,TW for the cosmochemical, geochemical and geodynamical estimates, respectively. Including the uncertainties, the predicted radiogenic heat production in BSE varies by a factor of four.

The bulk composition of the crust is relatively well defined. Our crustal model is constructed using CRUST2.0 \citep{bassin:2000} crustal structure including the densities of the layers. We treat the `A' and `B' type tiles of the CRUST2.0 model as oceanic and all other tiles as continental. We use HPE abundance estimates of \citet{rudnick:2003tgc} ``\RG'' for the continental crust and sediments. Abundances of the HPE in the oceanic crust are taken from \citet{white:2013tgc} ``\WK'', and for oceanic sediments we use those of \citet{plank:2013tgc} ``Plank''. \new{Within each crustal type, the uncertainties in U, Th, and K abundances are fully correlated. The uncertainties are uncorrelated between different crustal types.} Consequently, the continental crust (CC) generates $7.8\pm0.9$\,TW of radiogenic power, whereas the oceanic crust (OC) only gives off $0.22\pm0.03$\,TW (Table~\ref{tab:abund}). 

\section{Geoneutrino emission from Earth's mantle}
\label{sec:flux}

\subsection{Isochemical mantle models}
\label{sec:isochem}

From the radiogenic heat production in the BSE and the crust we calculate bulk mantle (BM) composition by a simple mass balance, 
\begin{equation} \label{abse}
  A_X^{BSE} m^{BSE} = A_X^{BM} m^{BM} + A_X^{CC} m^{CC} + A_X^{OC} m^{OC},
\end{equation}
where $A_X^Y$ is the elemental abundance of element $X$ in reservoir $Y$, and $m^Y$ is the mass of the reservoir (Table~\ref{tab:mass}). Equation \eqref{abse} assumes negligible radioactivity in the core \citep{mcdonough:2003tgc}. The input elemental abundances for BSE, CC, and OC, the reservoir masses, and BM abundances are listed in Table~\ref{tab:abund}. The resulting mantle Urey ratios amount to \new{$0.08\pm0.05$}, $0.3\pm0.1$ and $0.7\pm0.1$ for the cosmochemical, geochemical, and geodynamical BSE estimates. 
\new{The error in the Urey ratio arises from the errors in the surface heat flux (5\%), the crustal heat production (11\%), and the BSE heat production (10--18\%, depending on which estimate is used).}
The mantle radiogenic heat production can be as low as \new{1.3\,TW} (low-end cosmochemical BSE) and as high as 28\,TW (high-end geodynamical BSE), that is, a variation by a factor of \new{more than 20}.

We use the bulk mantle HPE abundances to predict geoneutrino fluxes at Earth's surface from a spherical-shell mantle of uniform composition (model UNIF in Figure~\ref{figcart}a). We account for the density increase with depth by a factor of roughly two across the mantle using PREM \citep{dziewonski:1981}, which also gives the radii of the surface (6371\,km), the top of the mantle (6346.6\,km) and the CMB (3480\,km). Here we neglect the variation in the crust--mantle boundary depth; however, the MOHO topography is accounted for later when we combine the fluxes from the mantle with the crustal flux.
The calculated mantle geoneutrino fluxes from $\Unum$ + $\Tnum$, in cm$^{-2}$\,$\mu$s$^{-1}$, are $0.28\pm0.19$, $1.0\pm0.3$, and $2.4\pm0.3$ for the three BSE estimates (black symbols in Figure~\ref{figcart}b).
\new{Geoneutrino fluxes from $\Unum$, $\Tnum$, and $\Knum$ are listed in Table~\ref{tab:flux}. Fluxes from $\Ufive$ scale with $\Unum$ fluxes in all models but are much smaller, $\frac{\Phi_{235}}{\Phi_{238}} = \frac{X_{235}}{X_{238}} \frac{\lambda_{235}}{\lambda_{238}} \frac{n_{235}}{n_{238}} = 0.0307$, largely due to the small $\Ufive/\Unum$ natural isotopic ratio. Uncertainties in fluxes reflect the uncertainties in abundance estimates, whereas the atomic parameter ($\lambda_X$, $M_X$, $\langle P \rangle$, $n_X$) uncertainties are negligible.}

\subsection{Layered mantle models}
\label{sec:layered}

A chemically uniform mantle with either geochemical or geodynamical HPE abundances is at odds with analyses of MORB sample compositions, which commonly require a MORB source rock depleted in HPEs relative to a bulk mantle. We consider several available compositional estimates for the depleted MORB-source mantle, as given by \citet{workman:2005} ``\WH'', \citet{salters:2004} ``\SaS'', and \citet{arevalo:2010} ``\AMcD'', listed here from the ``coldest'' (most depleted in HPEs) to the ``warmest'' compositions (Table~\ref{tab:abund}). The DM model of {\AMcD} is based on a global MORB composition and it deviates from the modeling used in \citet{arevalo:2009}, where they estimated the composition of the DM using differing proportions of N-MORB and E-MORB.

We consider two mantle reservoirs \new{with uniform composition}: a depleted mantle with DM composition above and enriched mantle (EM) below, where the reservoir masses satisfy $m^{BM} = m^{DM} + m^{EM}$. The elemental mass balance is then
\begin{equation} \label{amantle}
\new{
  A_X^{BM} = (1-F^{EM}) A_X^{DM} + F^{EM} A_X^{EM},
}
\end{equation}
\new{where we defined the mass fraction of the enriched reservoir, $F^{EM} = m^{EM}/m^{BM}$. Introducing the enrichment factor $E_X = A_X^{EM}/A_X^{DM}$, equation \eqref{amantle} can be rewritten as}
\begin{equation} \label{enrich}
\new{
  \frac{A_X^{BM}}{A_X^{DM}} = 1+(E_X-1)F^{EM}.
}
\end{equation}
For given BM and DM compositional estimates, a trade-off exists between the enrichment and the mass fraction of the enriched mantle (EM) reservoir -- for a prescribed DM composition, a smaller enriched reservoir mass requires larger chemical enrichment to satisfy a specified bulk mantle composition. 

The size of the enriched geochemical reservoir is not well constrained, with model values spanning a few percent to a few tens of percent of mantle by mass. \new{For our reference cases, we consider an enriched reservoir containing 10\% of mantle mass, somewhat arbitrarily chosen given the lack of robust constraints. We address the effect of the enriched reservoir size on the mantle geoneutrino signal in section \ref{sec:dVs}. Enrichment factors $E$ for various combinations of BSE and DM estimates
are listed in Table~\ref{tab:flux} (see \ref{sec:algebra} for details of the calculation). We impose a constraint of $E_X\geq1$ (or $A_X^{EM} \geq A_X^{DM}$), so that the ``enriched reservoir'' cannot be depleted relative to ``depleted mantle''. This constraint comes into effect for the low abundance cosmochemical BSE based on enstatite chondrite chemistry, making the cosmochemical estimate consistent with the absence of an enriched reservoir. Cosmochemical bulk mantle is too depleted to be consistent with {\AMcD} DM estimate at $1\sigma$ uncertainty level. It is also deficient in uranium, thorium, and potassium when combined with {\SaS} DM abundances, however consistent with this DM estimate when the uncertainty in abundances is considered (Table~\ref{tab:flux}).}

\new{Using eqn.\,\ref{flux2b} in \ref{sec:algebra}}
we calculate geoneutrino fluxes from a spherically symmetric two-reservoir mantle where the reservoir potentially enriched in HPEs is a 427\,km thick layer immediately above CMB (model EL in Fig.\,\ref{figcart}a). The predicted fluxes, including uncertainties, are listed in Table~\ref{tab:flux} and plotted in Figure~\ref{figcart}b as red, green and blue symbols. 
Relative to a uniform HPE distribution, a decrease in flux of geoneutrinos results when HPEs are sequestered at the bottom of the mantle, i.e., further from the measurement location at the Earth's surface \citep{dye:2010}.

What is the maximum possible flux reduction by such sequestration for a given bulk mantle HPE abundances? Maximum flux $\Phi_{max}$ is obtained for uniformly distributed HPEs \new{with $A_X^{BM}$ abundances throughout the mantle} (no enrichment, $E=1$). We exclude the dynamically implausible arrangement, where the deep mantle would be depleted in HPEs relative to the overlying mantle. Minimum possible flux $\Phi_{min}$ would be obtained in the hypothetical case where all HPEs were sequestered near CMB and the remaining mantle were HPE-free (\new{$A_X^{DM}=0$}, maximum enrichment, $E\rightarrow\infty$). In between these limit values, with increasing enrichment factor $E$ the flux $\Phi$ decreases proportionally to the depletion of the upper mantle \new{($\propto A_X^{DM}/A_X^{BM}$)}. Using eqn. \eqref{enrich} we get
\begin{equation}
  \label{reduce}
  \Phi(E) = \Phi_{min} + \frac{\Phi_{max}-\Phi_{min}}{1+(E-1)F^{EM}}.
\end{equation}
It is instructive to plot the normalized flux $\Phi(E)/\Phi_{max}$, which shows the flux reduction relative to a mantle with uniform HPE distribution (Figure~\ref{figcart}c). The normalized minimum flux $\Phi_{min}/\Phi_{max}$ for the EL model (enriched layer of uniform thickness) can be obtained analytically for a uniform density mantle.
PREM density mantle requires a simple integration and $\Phi_{min}/\Phi_{max}$ is 0.76 for $F^{EM}$ of 10\,\%. The inset in Figure~\ref{figcart}c shows the relatively weak dependence of this flux reduction limit on the mass fraction of the enriched layer.

\subsection{Models using a seismically constrained mantle structure}
\label{sec:tomo}

To illustrate the effect of possible lateral variation in the enriched reservoir geometry (e.g., LLSVPs or piles), we first consider axially symmetric cases with either a single deep-mantle pile or two antipodal piles (models P1 and P2, Figure~\ref{figcart}a). 
Model P1 is an idealized single 1000-km thick ``pile'' with vertical sides and lateral extent 0--76$^\circ$, sitting on the CMB. Model P2 has two antipodal piles of thickness 1000\,km and lateral extent 0--52$^\circ$. The piles in both models contain 10\% of the mantle by mass. The predicted geoneutrino fluxes from the mantle vary along latitude (Figure~\ref{figcart}d and Table~\ref{tab:flux}). We used geochemical BSE and {\AMcD} DM abundances, which lead to enrichment in U and Th within the piles by a factor of 6.0 and 12, respectively. Both models generate a surface-averaged flux which is basically identical (larger by 1\%) to the flux in the spherically symmetric EL model with the same HPE abundances. Model P1 shows a flux variation of~$^{+31\%}_{-22\%}$ amplitude about the average value, and model P2 shows a somewhat smaller variation of~$^{+18\%}_{-10\%}$ amplitude about the surface average.
The significant spatial variation of geoneutrino fluxes from the mantle motivates more detailed models of mantle geoneutrino emission.

We examine an enriched reservoir geometry that is based on seismic images of the deep mantle. We use seismic tomography model S20RTS \citep{ritsema:1999} and consider a simple mapping from shear-wave speed $V_S$ to enriched reservoir shape: slow regions with $V_S$ anomaly below $-0.25\,\%$ relative to PREM \citep{dziewonski:1981} and which are deeper than 1500\,km are assigned as enriched material. The remaining volume is assumed to be depleted mantle (model TOMO; Figure~\ref{figcart}a). This parameterization gives an enriched reservoir containing 9.5\,\% of the mantle by mass (or 8.2\,\% by volume), while 90.5\,\% is depleted mantle, i.e., proportions very similar to the previously presented two-reservoir mantle models, thus resulting in similar enrichment. 

The calculated mantle geoneutrino fluxes from the U and Th decay chains  vary with geographical location; a global map for one particular case using geochemical BSE and {\AMcD} DM abundances is shown in Figure~\ref{fig2D}. The surface-averaged flux is very close to the spherically symmetric EL model value (2\% larger;  Table~\ref{tab:flux}). The amplitude of the flux variation is~$^{+25\%}_{-13\%}$ relative to the spatial mean of $\new{0.96}\,\cmms$ (\Unum+\Tnum) for the enrichment factors 6.3 and 12 for U and Th, respectively. Two flux maxima -- one at 125\% of average signal in southwestern Africa ($9^\circ$\,S $13^\circ$\,E), the other at 121\% in Central Pacific ($9^\circ$\,S $161^\circ$\,W) -- are related to the African and Pacific deep mantle piles. The surrounding low flux region is broader and less pronounced. The absolute minimum at 87\% of the average is at $48^\circ$\,N $104^\circ$\,E (Mongolia). Mantle geoneutrino flux maps for all possible combinations of BSE and DM compositional estimates, including propagation of the uncertainties (Supplementary Figures~S1, S2) show that though the spatial pattern of the flux remains identical for all cases -- we use the same tomography-to-enriched reservoir mapping, the surface-averaged flux and the amplitude of variation is dependent on the compositional model. Table~\ref{tab:flux} reports the average, minimum, and maximum flux based on the central values of the compositional estimates. If the piles are compositionally distinct as indicated by geophysics, and correspond to enriched reservoirs as inferred from geochemistry, then the geoneutrino flux exhibits a dipolar pattern.

\subsubsection{Effect of mantle piles' size on geoneutrino flux}
\label{sec:dVs}

The size of the possible enriched reservoir is not well constrained from geochemical analyses. Seismic modeling defines the chemical piles beneath Africa at $\sim5\times10^9$\,km$^3$ (or $\sim0.6\,\%$ volume) \citep{wangwen:2004} and a similar size beneath the Pacific. This volume is smaller than the enriched volume fraction of 8\,\% (mass fraction of 10\,\%) we obtained by using a cut-off contour of $\delta V_s=-0.25\%$ of seismic model S20RTS \citep{ritsema:1999} to trace the enriched mantle reservoir boundary. We investigate how the mantle geoneutrino flux at Earth's surface changes when different $\delta V_s$ cut-off contours are used. More negative $\delta V_s$ cut-off results in a smaller enriched reservoir size, while the enrichment factor $E$ (relative to depleted mantle composition) is larger in order to yield a given bulk mantle composition (Table~\ref{tab:dVs}). Maps of mantle geoneutrino flux at the surface calculated for several different choices of $\delta V_s$ cut-off contours are shown in Figure~\ref{figdVs}. \new{As a result of the trade-off between the enriched reservoir size and its enrichment,} they show similar spatial pattern and comparable amplitudes.

\section{Is lateral variation in mantle geoneutrino flux resolvable?}
\label{sec:detect}

Measurements of geologically interesting electron antineutrinos include the detections of mantle ($M$) and crust ($C$) geoneutrinos, reactor ($r$) antineutrinos, and other antineutrino background ($bg$). The total event rate $R$ (in TNU) is
\begin{equation} \label{Rtot}
R = R_M + R_C + R_r + R_{bg}.
\end{equation}
After detector exposure $\varepsilon$ (in TNU$^{-1}$ or $10^{32}$\,proton\,yr) the expected total antineutrino count $N$ is
\begin{equation} \label{Ntot}
N = \varepsilon R.
\end{equation}
The exposure $\varepsilon$ is calculated from the detector of size $P$ (in units of $10^{32}$ free protons), detection efficiency $e$ ($0<e\leq1$) and live-time $T$ (in yr),
\begin{equation} \label{expo}
\varepsilon = e P T.
\end{equation}
A 10-kiloton detector contains about $8\times10^{32}$ free protons, therefore a year-long operation gives an exposure of $\sim8$\,TNU$^{-1}$ (assuming 100\,\% detection efficiency).

The detection count has a statistical error 
\begin{equation} \label{dNstat}
\delta N_\text{stat}=\sqrt{N}.
\end{equation}
Systematic errors come from instrumental error (in particular uncertainty $\delta\varepsilon$ in detector exposure), and the uncertainties in geological, reactor, and background signals. The uncertainty in mantle geoneutrino detection $\delta R_M$, written in terms of event rates and theirs errors, is obtained from \citep{dye:2010}
\begin{equation} \label{dRm}
(\delta R_M)^2 = \frac{R}{\varepsilon} + \left(\frac{R}{\varepsilon}\right)^2 (\delta\varepsilon)^2 + (\delta R_C)^2 + (\delta R_r)^2 + (\delta R_{bg})^2,
\end{equation}
where the first term on the right is the statistical error, followed by contributions to the systematic error: exposure, crust, reactor, background. 

Mantle geoneutrino determination at existing and proposed continental detection sites is limited by the uncertainty in crustal radioactivity. The dominance of crustal signature is clearly visible in Figure~\ref{figmer}, which maps total geoneutrino signal from crust + mantle (Figure~\ref{figmer}a), and the fraction of the signal that is contributed by the mantle (Figure~\ref{figmer}b).
\new{Predicted mantle event rates at existing detector sites are reported in Table~\ref{tab:sites}.}
In these calculations, MOHO topography is accounted for and the geoneutrino fluxes are evaluated at zero elevation in oceanic areas and at the Earth's surface (positive elevation) in continental regions.

Inspection of Figure~\ref{figmer}b suggests that the Pacific ocean basin offers the highest mantle-to-crust geoneutrino flux ratio. In Figure~\ref{figmer}c we show the variation of the predicted geoneutrino signal along the meridian at $161^\circ$W which intersects the Pacific mantle flux maximum at $9^\circ$S. \new{The crustal flux remains low between $35^\circ$N and $60^\circ$S at 2.0--4.0\,TNU (including uncertainty), while emission from the mantle varies between 2.4 and 30\,TNU depending on mantle compositional model and measurement location.} Mantle composition based on geodynamical BSE estimate results in highest geoneutrino fluxes and strongest spatial variation, a cosmochemical mantle model generates a small spatially uniform flux, and mantle based on geochemical BSE abundances is intermediate between the two. Importantly, with uncertainties considered, the three BSE estimates result in distinct mantle geoneutrino predictions at $1\sigma$ level (Figures~\ref{figmer}c and \ref{figr1r2}). 

In line with the general suggestion of \citet{dye:2010}, we propose that geoneutrino detection at two sites in the Pacific ocean is the best shot at constraining mantle U and Th abundances, and examining the thermochemical piles (superplumes) hypothesis. Site \#1 should be the location of the predicted Pacific mantle flux maximum ($161^\circ$W~$9^\circ$S, Figure~\ref{figmer}). Site \#2 should be remote from site \#1 so that the predicted mantle flux variation can be pronounced, while also sufficiently distant from continental crust in order to keep a favorable mantle-to-crust flux ratio; a good candidate is Southern Pacific (e.g., $161^\circ$W~$60^\circ$S, some 50 degrees directly south of site \#1, Figure~\ref{figmer}). 
The inputs for calculation of detection uncertainty $\delta R_M$ (eqn.\,\ref{dRm}) at each measurement site are $R_C\pm\delta R_C$, $R_r\pm\delta R_r$, $R_{bg}\pm\delta R_{bg}$, $\delta\varepsilon$, and $R_M$ \new{(Table~\ref{tab:sites})}. \new{We use exposure uncertainty of $\delta\varepsilon=2\,\%$ \citep{dye:2010}.}
A reasonable estimate for reactor background uncertainty $\delta R_r$ is $\pm5\,\%$ -- the uncertainties in the spectrum and cross section contribute $\sim2\,\%$, and further uncertainty is associated with power records from reactors, the oscillation parameters, and the reactor antineutrino anomaly \citep{dye:2012rg}.
Other background $R_{bg}$ consists of four primary sources: $^{13}\text{C}(\alpha,n)^{16}\text{O}$ reaction, fast neutrons from cosmic muons outside detector, long-lived neutron unstable radionuclides ($^9$Li, $^8$He) cosmogenically produced inside the detector, and accidentals. Borexino team estimated the background signal at $2.3\pm0.3$\,TNU \citep{bellini:2010} and we use this value as a conservative estimate; see more detailed discussion by \citet{dye:2012rg}.

Figure~\ref{figr1r2}a shows the predicted geoneutrino signal at proposed site \#1 at the Pacific flux maximum plotted against that at site \#2 in Southern Pacific. The detector exposure, necessary to discriminate between the predicted lateral variation in flux and a spherically uniform mantle emission, depends on the unknown mantle HPE abundances. The region of resolvable difference between predictions from a ``piles'' model and from a uniform mantle model is highlighted in Figure~\ref{figr1r2}a. Exposure $\lesssim10$\,TNU$^{-1}$ is sufficient to resolve the variation predicted from geodynamical BSE models at $1\sigma$ uncertainty level. The lateral variation is resolvable for the high-abundance end of the geochemical BSE model with exposures from $\sim10$ to few tens TNU$^{-1}$ (Figure~\ref{figr1r2}b). Cosmochemical predictions and the low end of geochemical predictions produce a mantle essentially uniform in composition.

\section{Discussion}
\label{sec:discussion}

Combined analysis of KamLAND \citep{araki:2005nat,gando:2011bis} and Borexino \citep{bellini:2010} electron antineutrino observations places the bounds on mantle geoneutrino event rate at $23\pm10$\,TNU where the Th/U ratio spans a range of 2.7--3.9 \citep{fiorentini:2012prd}. 
This is a result with a relatively large error, which supports both geodynamical and geochemical BSE models, but is incompatible with cosmochemical BSE \citep[and collisional erosion models such as][]{oneill:2008} at $1\sigma$ level (Figure~\ref{figcart}b).
KamLAND now benefits from significant decrease of nuclear reactor signal after power plant shutdowns following the Fukushima Daiichi accident. Borexino's result is dominated by statistical uncertainty, which decreases with continuing measurement.
New experiments capable of geoneutrino detection are being developed. In 2013 the SNO+ detector at SNOLab in Ontario, Canada, is expected to go on-line. The LENA experiment  is proposed either at the Pyh\"asalmi mine (near Pyh\"aj\"arvi, Finland), or at the Laboratoire Souterrain de Modane \citep[near Fr\'ejus, France;][]{LENA:pub}. 
Reduction of instrumental uncertainty and more precise description of crustal geology, particularly in the vicinity of neutrino experiment sites, are expected increase sensitivity to the distribution of Earth's internal radioactivity.

The debate about the chemical composition of the silicate Earth remains open. \new{Latest studies find support for both enstatite chondrite-derived composition \citep{warren:2011,zhangdauphas:2012} and carbonaceous chondrite-based composition \citep{murakami:2012}, some propose a more complicated chondrite mix \citep{fitoussi:2012}, or argue against a chondritic Earth altogether \citep{campbell:2012}.}
Geoneutrinos can supply the key evidence necessary to refine our knowledge of Earth's heat engine. If BSE abundances turn out to be close to the low cosmochemical estimate, for example, geophysics will be challenged to explain the present-day high surface heat flux. Detection of lateral variation in the mantle geoneutrino flux -- or absence thereof -- will stimulate further well-posed questions about the stability and dynamics of the chemical piles, and the origin and nature of the seismically imaged deep-mantle structures. These questions clearly motivate experimental efforts to constrain mantle radioactivity by geoneutrino detection.

Our findings highlight the potential for doing neutrino tomography of the mantle. From the perspective of deep-Earth research, the desired location for a geoneutrino detector is an oceanic site far away from continental crust; an oceanic transportable detector is proposed for the Hanohano experiment \citep{learned:2008arxiv}. Geoneutrino detection at two sites in the Pacific ocean offers a possibility to constrain mantle uranium and thorium abundances, and to examine the thermochemical piles hypothesis.
In general, adding an observation datum with a reasonably low uncertainty ($\lesssim15\,\%$) to Figure~\ref{figr1r2} would substantially tighten the constraints on mantle radioactivity abundance and distribution. Contingent on enthusiastic involvement of the geophysical community, experimental neutrino research can contribute significantly to our understanding of Earth's interior.

\section*{Acknowledgments}
\new{We wish to thank Fabio Mantovani and an anonymous reviewer for their detailed and thoughtful reviews.} We gratefully acknowledge support for this research from NSF EAR 0855791 CSEDI Collaborative Research: Neutrino Geophysics: Collaboration Between Geology \& Particle Physics, and Hawaii Pacific University's Trustees' Scholarly Endeavors Program.

\appendix 

\new{
\section{Antineutrino flux algebra}
\label{sec:algebra}

Antineutrino flux $\Phi$ from a geological reservoir $\Omega$ of uniform compositional abundances $A_X$ is calculated, using eqns.\,\eqref{flux0} and \eqref{abundance}, as
\begin{equation} \label{flux1}
  \Phi_X(\rr) = P_X A_X G^\Omega(\rr),
\end{equation}
where the prefactor $P_X=n_X \lambda_X X_X \langle P \rangle/M_X$ contains the atomic parameters, and the geological response factor $G^\Omega$, defined as
\begin{equation} \label{G^{BM}}
  G^\Omega(\rr) = \frac{1}{4\pi} \int\limits_\Omega \frac{\rho(\rr')}{|\rr-\rr'|^2} \mathrm{d}\rr'
\end{equation}
\citep[e.g.,][]{dye:2012rg}, depends on the geometry and density structure of the reservoir. 

Geoneutrino flux from a two-reservoir mantle (DM+EM) is readily calculated as (hereafter dropping the $\rr$-dependence of $\Phi$ and $G$)
\begin{equation} \label{flux2a}
  \Phi_X = P_X \left(A_X^{DM} G^{DM} + A_X^{EM} G^{EM}\right),
\end{equation}
Using eqn.\,\eqref{amantle} and noting that $G^{DM}=G^{BM}-G^{EM}$ we can rewrite the flux \eqref{flux2a} as a linear combination of bulk mantle and depleted mantle abundances,
\begin{equation} \label{flux2b}
  \Phi_X = P_X \left[A_X^{DM} \left(G^{BM}-\frac{G^{EM}}{F^{EM}}\right) + A_X^{BM} \frac{G^{EM}}{F^{EM}}\right].
\end{equation}
This equation allows a straightforward {\em exact} calculation of the flux $\Phi_X$ {\em and} its uncertainty for a given reservoir structure ($G^{BM}$, $G^{EM}$, $F^{EM}$) as the uncertainty of the atomic parameters ($P_X$) is negligible relative to uncertainty in abundances ($A_X^{BM}$, $A_X^{EM}$). The enrichment factors $E_X=A_X^{EM}/A_X^{DM}$ are then calculated from
\begin{equation} \label{enrich2}
  E_X = 1 + \frac{\Phi_X-\Phi_X^{llim}}{P_X A_X^{DM} G^{EM}},
\end{equation}
where $\Phi_X^{llim} = P_X A_X^{DM} G^{BM}$ is the lower limit on flux emitted from a uniform mantle with depleted mantle composition. The constraints of $E_X\geq1$ and the equivalent constraints of $\Phi_X \geq \Phi_X^{llim}$ are applied a posteriori.
}


\renewcommand\thetable{\arabic{table}}
\renewcommand\thefigure{\arabic{figure}}


\clearpage

\begin{table}
\caption{Atomic parameters. Atomic mass $M$ in unified atomic mass units ($1u=1.661\times10^{-27}$\,kg), half-life $\tau_{1/2}$ in Gyr, decay constant $\lambda$ in $10^{-18}$\,s$^{-1}$, energy available for radiogenic heating $Q_h$ in pJ per decay. $^\dagger$Non-integer $\eanu$'s per chain value for $^{40}$K reflects branching into $\beta$ decay and electron capture.}
\label{tab:atomic}
\vspace{1em}
\begin{tabular}{l|lllll}
\hline 
 & $^{238}$U & $^{235}$U & $^{232}$Th & $^{40}$K & Reference \\
\hline 
Isotopic abundance $X$ & 0.9927 & 0.007204 & 1.0000 & 117\,ppm & www.nist.gov \\
Atomic mass $M$ & 238.051 & 235.044 & 232.038 & 39.9640 & www.nist.gov \\
Half life $\tau_{1/2}$ &  4.468 & 0.704 & 14.05 & 1.265 & www.nucleide.org \\
Decay constant $\lambda$ & 4.916 & 31.2 & 1.563 & 17.36 & $\lambda=\ln(2)/\tau_{1/2}$ \\
Energy to heat $Q_h$ & 7.648 & 7.108 & 6.475 & 0.110 & \citet{dye:2012rg} \\
$\eanu$'s per chain $n$ & 6 & 4 & 4 & $0.8928^\dagger$ &  \\
\hline
\end{tabular}
\end{table}

\begin{table}
\small 
\caption{Compositional estimates for heat producing elements (HPEs), corresponding radiogenic power, and the mantle Urey ratio. Bulk silicate Earth (BSE): cosmochemical, geochemical, and geodynamical estimates (see text). Bulk continental crust (CC, includes sediments): \RG\ \citep{rudnick:2003tgc}. Bulk oceanic crust (OC, includes sediments): \WK\ \citep{white:2013tgc}, Plank \citep{plank:2013tgc}. Bulk mantle (BM) calculated from eqn.\,\ref{abse}. Depleted Mantle (DM), MORB-source: \WH\ \citep{workman:2005}, \SaS\ \citep{salters:2004}, \AMcD\ \citep{arevalo:2010}. $^*$Assumes that entire mantle is DM.}
\label{tab:abund}
\vspace{1em}
\begin{tabular}{l|lll|l|l}
\hline
& \multicolumn{3}{c|}{\bf BSE} & {\bf CC (incl.\,sed.)} & {\bf OC (incl.\,sed.)} \\
\hline
& Cosmochem.  & Geochem. & Geodyn. & \RG & \WK, Plank \\
\hline
$A_U$ in ppb & $12\pm2$ & $20\pm4$ & $35\pm4$ & $1.47\pm0.25$\,ppm & $0.15\pm0.02$\,ppm \\
$A_{Th}$ in ppb & $43\pm4$ & $80\pm13$ & $140\pm14$ & $6.33\pm0.50$\,ppm & $0.58\pm0.07$\,ppm \\
$A_K$ in ppm & $146\pm29$ & $280\pm60$ & $350\pm35$ & $1.63\pm0.12$\,wt\% & $0.16\pm0.02$\,wt\% \\
Th/U & 3.5 & 4.0 & 4.0 & 4.3 & 3.9 \\
K/U & 12000 & 14000 & 10000 & 11100 & 10400 \\
Power in TW & $11\pm2$ & $20\pm4$ & $33\pm3$ & $7.8\pm0.9$ & $0.22\pm0.03$ \\
\hline
\end{tabular}
\par
\vspace{1em}
\begin{tabular}{l|lll|lll}
\hline
& \multicolumn{3}{c|}{\bf BM} & \multicolumn{3}{c}{\bf DM} \\
\hline
& Cosmochem.  & Geochem. & Geodyn. & \WH & \SaS & \AMcD \\
\hline
$A_U$ in ppb & $4.1\pm2.8$ & $12\pm4$ & $27\pm4$ & $3.2\pm0.5$ & $4.7\pm1.4$ & $8\pm2$ \\
$A_{Th}$ in ppb & $8.4\pm5.1$ & $46\pm12$ & $106\pm14$ & $7.9\pm1.1$ & $13.7\pm4.1$ & $22\pm4$ \\
$A_K$ in ppm & $57\pm30$ & $192\pm61$ & $263\pm36$ & $50\pm8$ & $60\pm17$ & $152\pm30$ \\
Th/U & 2.0 & 3.8 & 3.9 & 2.5 & 2.9 & 2.8  \\
K/U & 13900 & 16000 & 9700 & 15600 & 12800 & 19000 \\
Power in TW & $3.3\pm2.0$ & $12\pm4$ & $25\pm3$ & $2.8\pm0.4^*$ & $4.1\pm1.2^*$ & $7.5\pm1.5^*$ \\
Mantle Urey ratio & $0.08\pm0.05$ & $0.3\pm0.1$ & $0.7\pm0.1$ & \\
\hline
\end{tabular}
\end{table}

\begin{table}
\caption{(a) Earth reservoir masses. Values from PREM \citep{dziewonski:1981} and CRUST2.0 \citep{bassin:2000}.}
\label{tab:mass}
\vspace{1em}
\begin{tabular}{llrl}
\hline
Reservoir & & Mass in kg & Reference \\
\hline
Earth & $m_E$ & $5.9732\times10^{24}$ & PREM \\
Continental crust (incl.\,sed.) & $m_{CC}$ & $2.14\times10^{22}$ & CRUST2.0 \\
Oceanic crust (incl.\,sed.) & $m_{OC}$ & $0.63\times10^{22}$ & CRUST2.0 \\
Crust (=cont.+oc.) & $m_C$ & $2.77\times10^{22}$ & \\
Mantle & $m_M$ & $4.0024\times10^{24}$ & PREM \\
BSE (=mantle+crust) & $m_{BSE}$ & $4.0301\times10^{24}$ & \\
\hline
\end{tabular}
\end{table}

\renewcommand{\baselinestretch}{1.1}

\begin{table}
\small 
\caption{Enrichment factors and geoneutrino fluxes from the mantle for various models of HPE abundances and distribution. Results for spherically symmetric models (UNIF, EL) are reported including $1\sigma$ uncertainties. For models with lateral variation in HPE abundances (P1, P2, TOMO), the surface average, minimum and maximum flux values (\f{ave}{min}{max}) based on central value of compositional estimates are shown. \new{``n/a'' indicates inconsistency for the particular combination of BSE and DM compositional estimates (i.e., deficiency in HPE).}}
\label{tab:flux}
\vspace{1em}
\hspace{-7em}
\begin{tabular}{ll|lll|l|lll}
\hline
BSE & DM & \multicolumn{3}{c|}{Enrichment factor $E$} & Model & \multicolumn{3}{c}{Geoneutrino flux $\Phi$ in cm$^{-2}$\,$\mu$s$^{-1}$} \\
& & U & Th & K & & $\Unum$ & $\Tnum$ & $\Knum$ \\
\hline
\multicolumn{9}{l}{\bf Spherically symmetric models --- EM is 10\,\% of mantle by mass} \\
\hline
Cosmochem. & --- &  --- & --- & --- & UNIF & $0.20\pm0.13$ & $0.088\pm0.053$ & $0.98\pm0.52$ \\
Geochem. & --- &  --- & --- & --- & UNIF & $0.57\pm0.20$ & $0.48\pm0.13$ & $3.3\pm1.1$ \\
Geodyn. & --- &  --- & --- & --- & UNIF & $1.3\pm0.2$ & $1.1\pm0.1$ & $4.6\pm0.6$ \\
\hline
Cosmochem. & \AMcD & n/a & n/a & n/a & EL & --- & --- & --- \\
Cosmochem. & \SaS & 1--5.8 & 1 & 1--5.5 & EL & 0.22--0.30 & 0.14 & 1.0--1.4 \\
Cosmochem. & \WH & $\fpm{3.8}{-2.8}{+8.8}$ & $\fpm{1.7}{-0.7}{+6.4}$ & \fpm{2.3}{-1.3}{+6.0} & EL & \fpm{0.18}{-0.03}{+0.10} & \fpm{0.087}{-0.005}{+0.040} & \fpm{0.96}{-0.09}{+0.40} \\
\hline
Geochem. & \AMcD & $\fpm{6.0}{-5.0}{+5.3}$ & $12\pm6$ & $\fpm{3.6}{-2.6}{+4.0}$ & EL & \fpm{0.53}{-0.15}{+0.16} & $0.42\pm0.10$ & \fpm{3.2}{-0.5}{+0.8} \\
Geochem. & \SaS & $17\pm9$ & $24\pm9$ & $23\pm10$ & EL & $0.49\pm0.15$ & $0.40\pm0.10$ & $2.8\pm0.8$ \\
Geochem. & \WH & $29\pm13$ & $49\pm16$ & $29\pm12$ & EL & $0.47\pm0.15$ & $0.38\pm0.10$ & $2.7\pm0.8$ \\
\hline
Geodyn. & \AMcD & $25\pm5$ & $39\pm7$ & $8.3\pm2.4$ & EL & $1.1\pm0.1$ & $0.90\pm0.11$ & $4.1\pm0.5$ \\
Geodyn. & \SaS & $49\pm8$ & $69\pm11$ & $35\pm6$ & EL & $1.0\pm0.1$ & $0.88\pm0.11$ & $3.7\pm0.5$ \\
Geodyn. & \WH & $76\pm12$ & $126\pm18$ & $44\pm7$ & EL & $1.0\pm0.1$ & $0.86\pm0.11$ & $3.7\pm0.5$ \\
\hline
\multicolumn{9}{l}{\bf Laterally variable cartoon models --- EM is 10\,\% of mantle by mass} \\
\hline
Geochem. & \AMcD & 6.0 & 12 & 3.6 & P1 & \f{0.53}{0.44}{0.66} & \f{0.42}{0.31}{0.59} & \f{3.2}{2.9}{3.6} \\
Geochem. & \AMcD & 6.0 & 12 & 3.6 & P2 & \f{0.53}{0.49}{0.61} & \f{0.42}{0.37}{0.52} & \f{3.2}{3.0}{3.5} \\
\hline
\multicolumn{9}{l}{\bf Seismic tomography-based models --- EM is 9.5\,\% of mantle by mass} \\
\hline
Cosmochem. & \AMcD &  n/a & n/a & n/a &  & --- & --- & --- \\
Cosmochem. & \SaS &  n/a & n/a & n/a &  & --- & --- & --- \\
Cosmochem. & \WH & 4.0 & 1.8 & 2.4 & TOMO & \f{0.19}{0.17}{0.21} & \f{0.087}{0.085}{0.090} & \f{0.96}{0.93}{1.02} \\
\hline
Geochem. & \AMcD & 6.3 & 12 & 3.8 & TOMO & \f{0.53}{0.48}{0.64} & \f{0.43}{0.36}{0.57} & \f{3.2}{3.0}{3.5} \\
Geochem. & \SaS & 17 & 26 & 24 & TOMO & \f{0.50}{0.40}{0.70} & \f{0.41}{0.32}{0.60} & \f{2.9}{2.2}{4.2} \\
Geochem. & \WH & 30 & 51 & 31 & TOMO & \f{0.49}{0.37}{0.72} & \f{0.40}{0.29}{0.62} & \f{2.9}{2.1}{4.2} \\
\hline
Geodyn. & \AMcD & 26 & 41 &  8.6 & TOMO & \f{1.1}{0.9}{1.6} & \f{0.93}{0.68}{1.42} & \f{4.2}{3.6}{5.2} \\
Geodyn. & \SaS & 51 & 72 & 36 & TOMO & \f{1.1}{0.8}{1.7} & \f{0.92}{0.64}{1.45} & \f{3.9}{2.9}{5.8} \\
Geodyn. & \WH & 80 & 132 & 46 & TOMO & \f{1.1}{0.7}{1.7} & \f{0.90}{0.61}{1.47} & \f{3.8}{2.8}{5.9} \\
\hline
\end{tabular}
\end{table}

\renewcommand{\baselinestretch}{1.5}

\begin{table}
\caption{Mass fraction and enrichment factors for the enriched mantle reservoir obtained for various $\delta V_s$ cut-off contours in the TOMO model.}
\label{tab:dVs}
\vspace{1em}
\centering
\begin{tabular}{lllll}
\hline 
$\delta V_s$ cut-off & EM mass. frac. & \multicolumn{3}{c}{Enrichment factor} \\
 & $F^{EM}$ & $E_U$ & $E_{Th}$ & $E_K$ \\
\hline 
$-0.25\,\%$ & 9.5\,\% & 6.3 & 12 & 3.8 \\
$-0.50\,\%$ & 4.4\,\% & 13 & 26 & 7.0 \\
$-0.75\,\%$ & 1.8\,\% & 30 & 63 & 16 \\
$-1.00\,\%$ & 0.71\,\% & 72 & 155 & 38 \\
\hline
\end{tabular}
\end{table}

\begin{table}
\caption{\new{Predicted event rates from the mantle $R_M$ and the crust $R_C$ at at the sites of existing geoneutrino detectors and at the proposed locations of the two-site oceanic measurement. Reactor rates $R_r$ \citep[from][]{dye:2012rg} at the proposed sites used for calculation of the mantle rate detection uncertainty are also listed. Event rates in TNU. $^\dagger$Crustal rates from \citet{fiorentini:2012prd}.}}
\label{tab:sites}
\vspace{1em}
\hspace{-4em}
\begin{tabular}{llllllll}
\hline
Site & Lat. & Lon. & \multicolumn{3}{c}{Mantle event rate $R_M$} & $R_C$ & $R_r$ \\
 & $^\circ$N & $^\circ$E & Cosmochem. & Geochem. & Geodyn. & Crust & Reactor \\
\hline
Kamioka & 36.43 & 137.31 & 2.3--3.8 & \fpm{7.2}{-2.6}{+2.8} & \fpm{14.4}{-2.4}{+2.6} & $26.5\pm2.0^\dagger$ \\
Gran Sasso & 42.45 & 13.57 & 2.3--4.3 & \fpm{8.4}{-2.9}{+3.0} & \fpm{18.3}{-2.7}{+2.8} & $25.3\pm2.8^\dagger$ \\
Sudbury & 46.47 & $-81.20$ & 2.3--3.7 & \fpm{6.9}{-2.5}{+2.7} & \fpm{13.5}{-2.3}{+2.6} &  \\
Site \#1 & $-9$ & $-161$ & 2.4--5.4 & \fpm{10.7}{-4.1}{+4.3} & \fpm{25.5}{-4.1}{+4.0} & $2.5\pm0.3$ & $0.9\pm5\%$ \\
Site \#2 & $-60$ & $-161$ & 2.3--4.0 & \fpm{7.7}{-2.7}{+2.9} & \fpm{15.9}{-2.6}{+2.7} & $3.4\pm0.4$ & $0.6\pm5\%$ \\ 
\hline
\end{tabular}
\end{table}


\clearpage
\begin{figure}
\caption{(a) Cartoon model gallery. Bulk mantle in dark green, depleted mantle (DM) in light green, and enriched mantle (EM) in dark red. Models UNIF and EL are spherically symmetric, models P1 and P2 are axially symmetric. (b) Calculated geoneutrino fluxes from $\Unum$+$\Tnum$ decay in a spherically symmetric mantle (black, red, green, and blue data points and error bars) compared to observation \citep[orange region, combined analysis of KamLAND and Borexino data;][]{fiorentini:2012prd}. Conversion between $\cmms$ and TNU on right-hand vertical axis assumes Th/U=3.9. (c) Effect of HPE sequestration in a deep mantle layer on the geoneutrino flux at the surface. Main plot shows flux reduction with increasing enrichment of the deep-seated reservoir. Dependence of the maximum flux reduction on the enriched reservoir size is shown in the inset. (d) Mantle geoneutrino flux ($\Unum$+$\Tnum$) variation along latitude for cartoon models shown in panel `a' using geochemical BSE and {\AMcD} DM compositional estimates.}
\label{figcart}
\end{figure}
\includegraphics{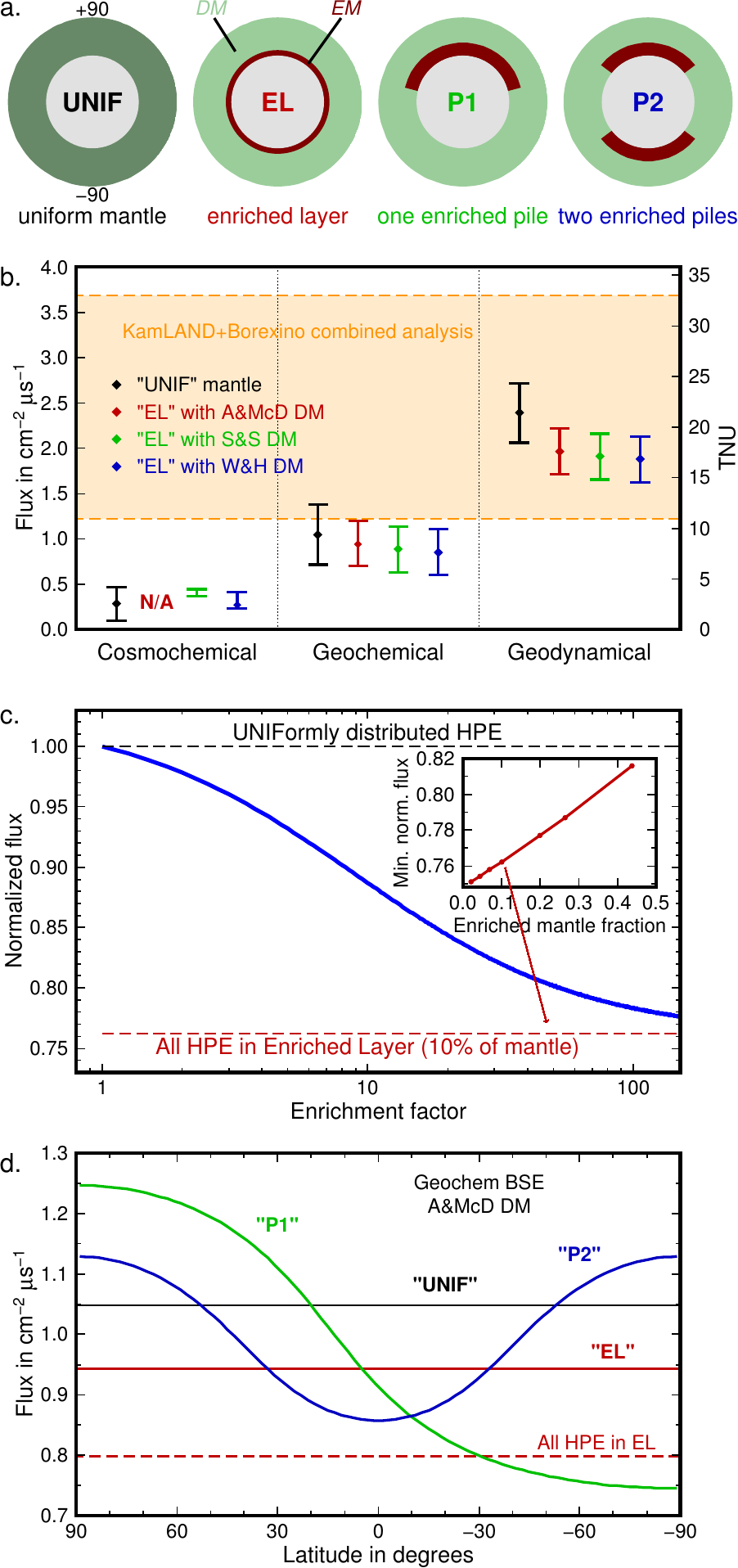}

\clearpage
\begin{figure}
\includegraphics{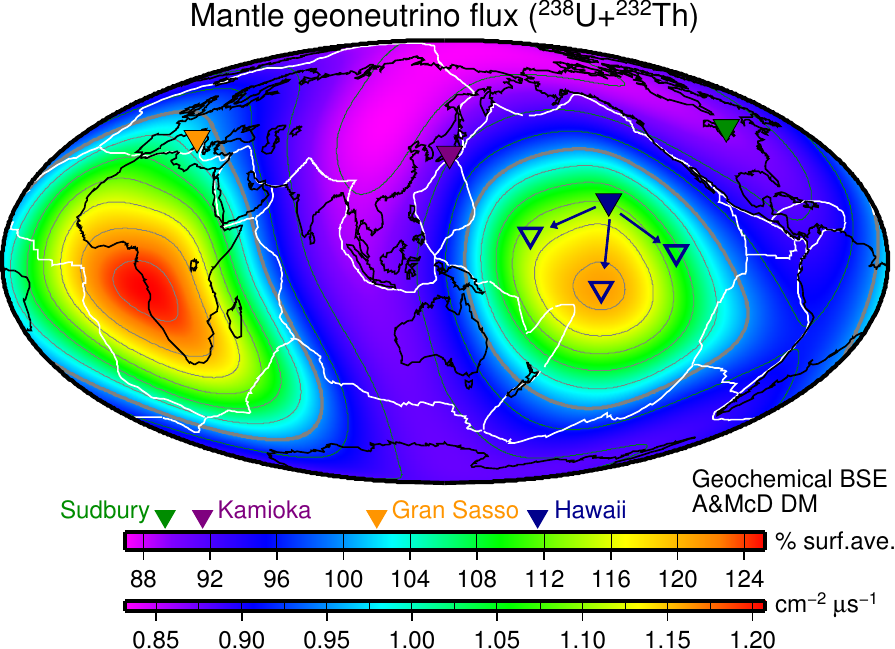}
\caption{Global map of geoneutrino flux from $\Unum$+$\Tnum$ decay in the mantle calculated for the TOMO model using geochemical BSE and {\AMcD} DM compositional estimates. A uniform radius for the crust--mantle boundary is used (6346.6\,km), flux is evaluated at radius of 6371\,km and shown as percentage of the surface-averaged value (color scale) with contour lines at 4\% intervals. Continental outlines (black), plate boundaries (white), and locations of geoneutrino detectors are plotted: Kamioka, Japan (KamLAND, operational); Gran Sasso, Italy (Borexino, operational); Sudbury, Canada (SNO+, online 2013); Hawaii (Hanohano, proposed; transportable detector as illustrated by open triangles and arrows).}
\label{fig2D}
\end{figure}

\clearpage
\begin{figure}
\includegraphics{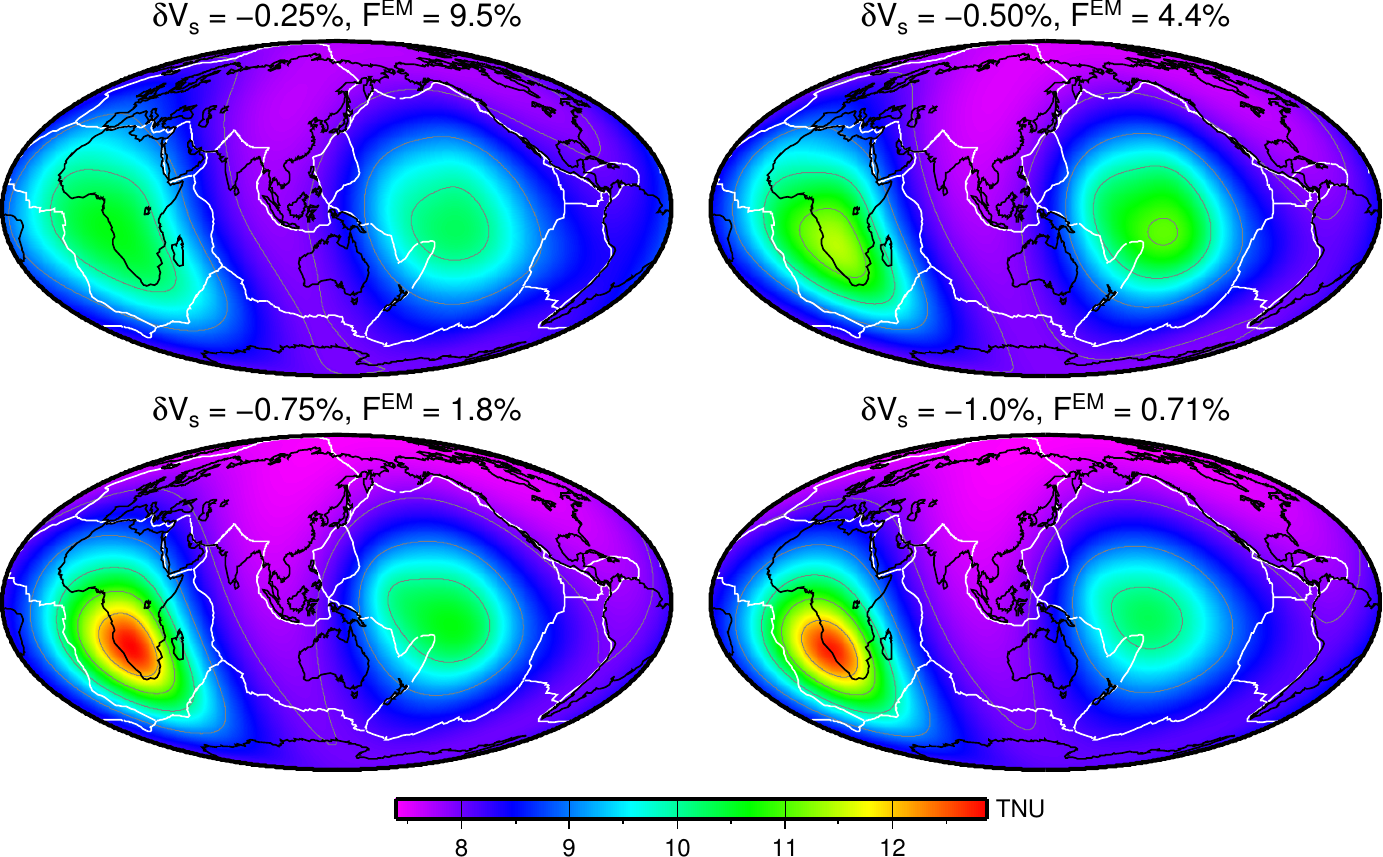}
\caption{Global map of geoneutrino event rate in TNU from $\Unum$+$\Tnum$ decay in the mantle calculated for the TOMO model using geochemical BSE and {\AMcD} DM compositional estimates, and several different cut-off $\delta V_S$ contours (indicated above each map together with the resulting mass fraction of the enriched mantle reservoir $F^{EM}$). A unique radius for the crust--mantle boundary is used (6346.6\,km), flux is evaluated at radius of 6371\,km and shown in TNU with contour lines at 1\,TNU intervals. Continental outlines (black) and plate boundaries (white) are plotted. Color scale is identical for all four maps.}
\label{figdVs}
\end{figure}

\clearpage
\begin{figure}
\caption{(a) Global map of predicted total geoneutrino signal ($\Unum$+$\Tnum$, crust+mantle) in TNU. Mantle anti-neutrino emission model same as in Fig.\,\ref{fig2D}. Crustal prediction based on CRUST2.0 structure, and {\RG}, {\WK}  and Plank compositional estimates (see text). Topography of the crust--mantle boundary is accounted for, geoneutrino fluxes are evaluated at zero elevation in oceanic areas and at Earth's surface in continental regions. Continental outlines (black) and plate boundaries (white) are shown.
(b) Map showing the fraction of total signal from panel `a' that is contributed by the mantle; the remainder is the crustal contribution. Contour lines at 10\% intervals. (c) Variation of predicted geoneutrino signal along $161^\circ$W meridian which intersects the Pacific mantle flux maximum at $9^\circ$S. Crustal prediction shown in brown. Mantle predictions based on cosmochemical, geochemical, and geodynamical BSE estimates shown in blue, green, and red, respectively. Central values (thick curves) and $1\sigma$ uncertainty limits (thin lines and shading) are shown. Two oceanic measurement sites are proposed (shown in panels `b' and `c') in order to constrain Earth's mantle architecture.}
\label{figmer}
\end{figure}
\includegraphics{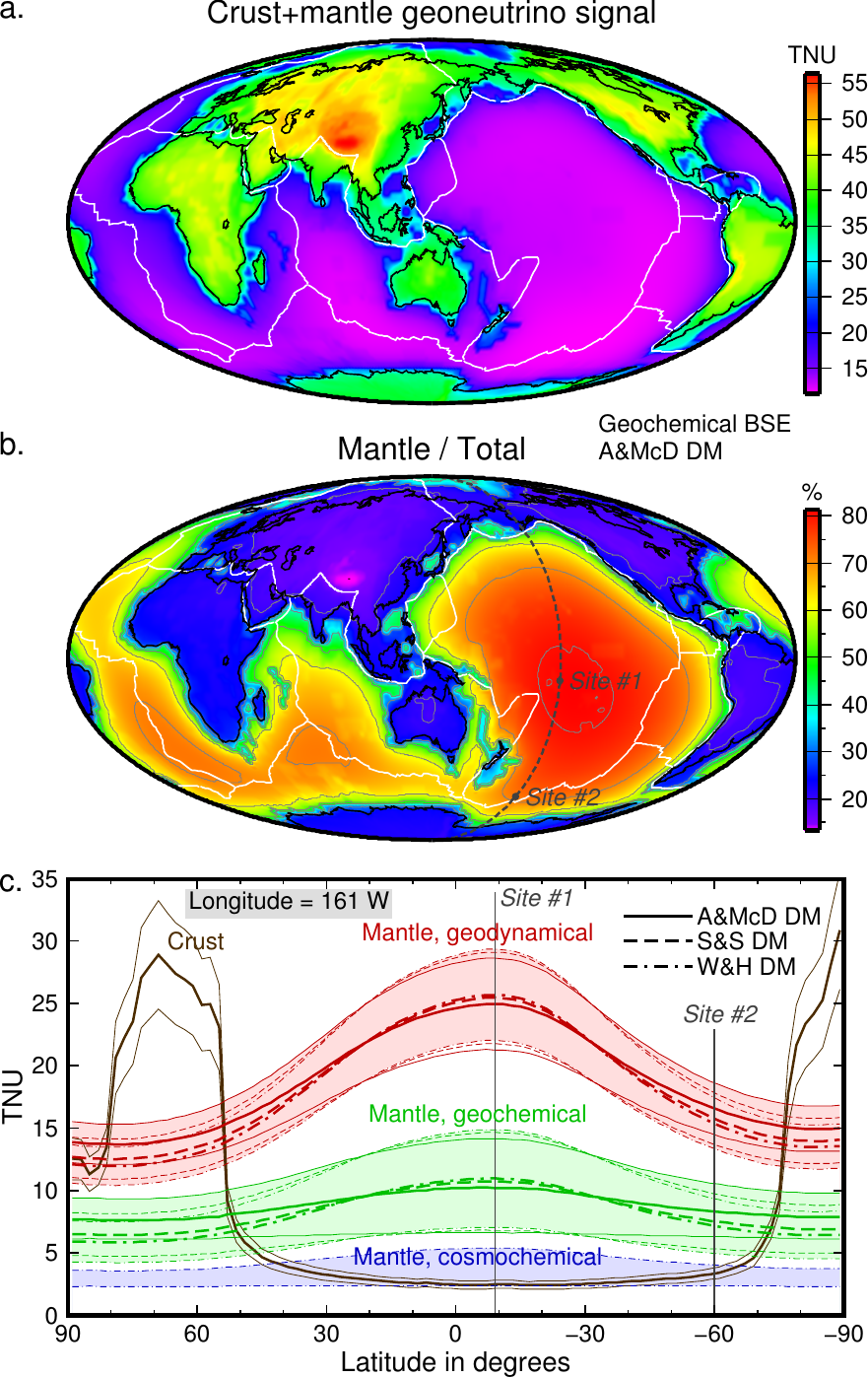}

\clearpage
\begin{figure}
\includegraphics{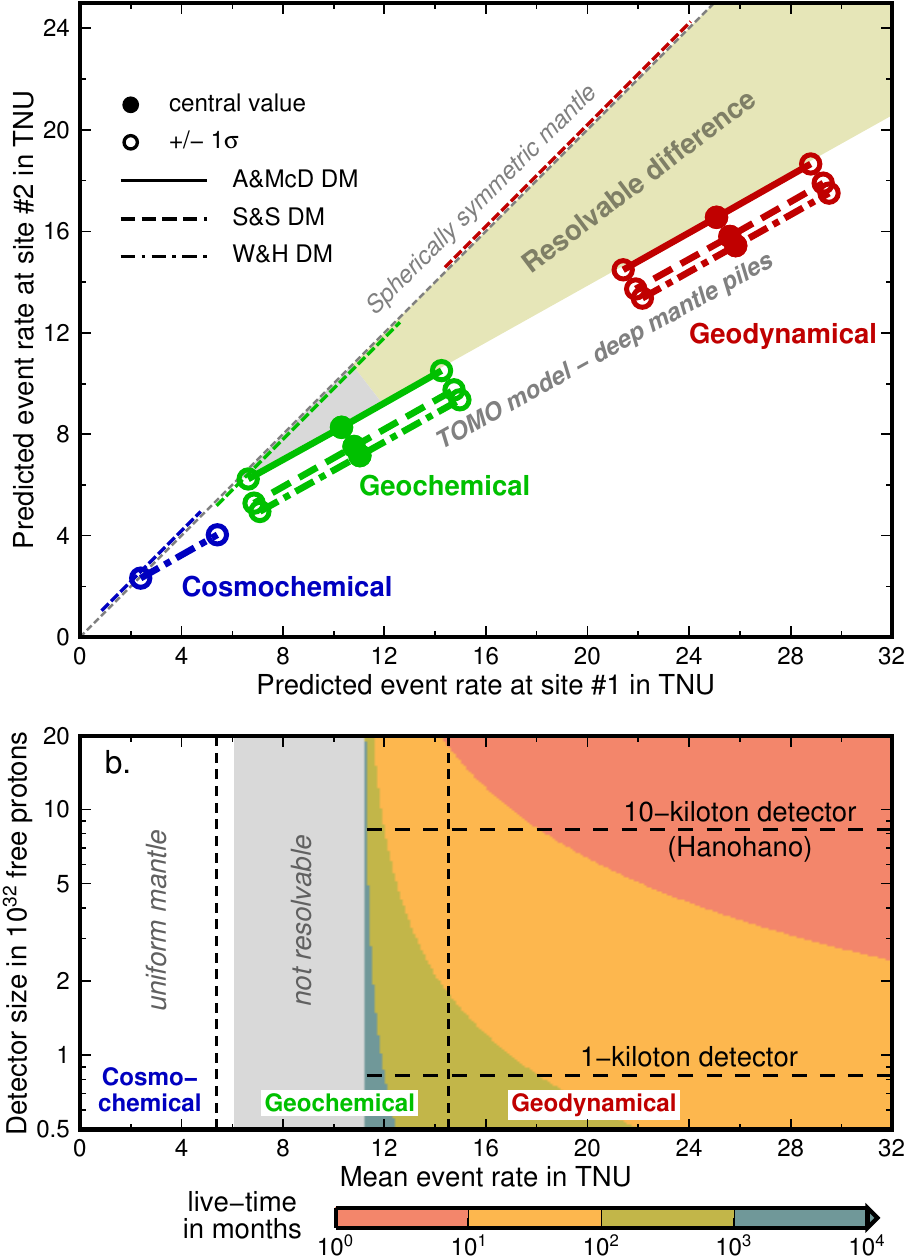}
\caption{(a) Mantle geoneutrino event rate in TNU at site \#1 (horizontal axis) versus that at site \#2 (vertical axis) as predicted from cosmochemical (blue), geochemical (green) and geodynamical (red) BSE estimates, including $1\sigma$ uncertainties, using various DM estimates. Predictions for a spherically symmetric mantle, both homogeneous and layered, follow a straight line with slope 1. Predictions of TOMO model align along a gentler slope. The region of resolvable difference between these two predictions is indicated. (b) The detector exposure required to discriminate between the two predictions shown in terms of detector size (vertical axis) and live-time (color) as a function of mean event rate at sites \#1 and \#2 (horizontal axis). Vertical dashed lines separate regions of different BSE estimates.}
\label{figr1r2}
\end{figure}


\renewcommand{\thefigure}{S\arabic{figure}}
\setcounter{figure}{0}

\clearpage

\noindent {\bf \large Supplemetary Figures}

\begin{figure}[h]
\caption{Global map of geoneutrino flux from $\Unum$+$\Tnum$ decay in the mantle calculated for the TOMO model for all combinations of BSE and DM compositional estimates. A unique radius for the crust--mantle boundary is used (6346.6\,km), flux is evaluated at radius of 6371\,km and shown in TNU. Continental outlines (black) and plate boundaries (white) are plotted. Middle column calculated using central values of the enrichment factor, right and left columns calculated at $\pm1\sigma$ limits for the enrichment factor in each case. The color scale is common within each BSE estimate used. Empty maps labeled ``n/a'' reflect inconsistency for the particular combination of BSE and DM compositional estimates (enrichment factor smaller than one or even negative).}
\label{figall1}
\end{figure}

\begin{figure}[h]
\caption{Same as Figure \ref{figall1} except for the color scale, which is identical for all maps.}
\label{figall2}
\end{figure}

\pagestyle{empty}

\includegraphics[scale=0.9]{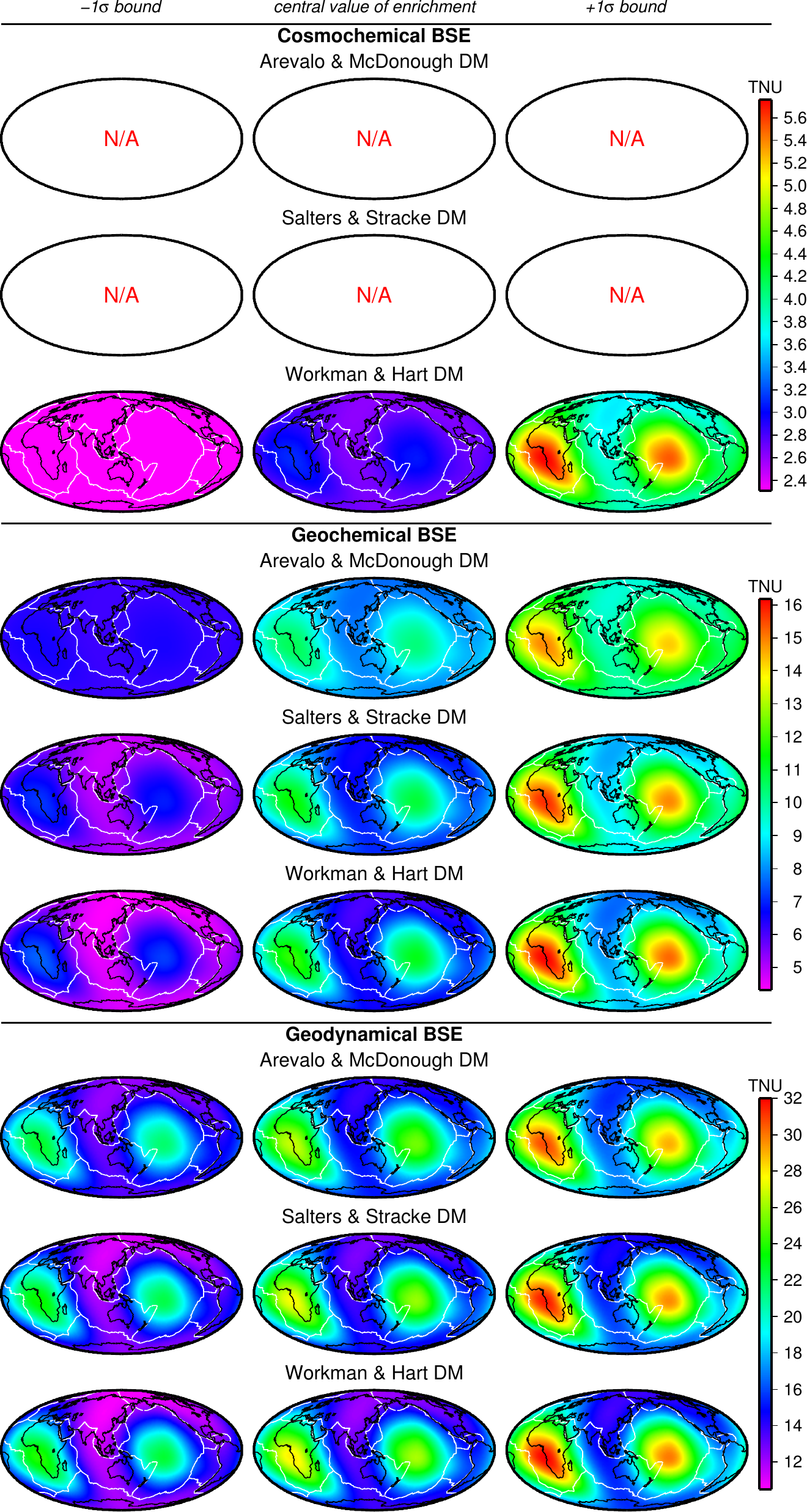}

\includegraphics[scale=0.9]{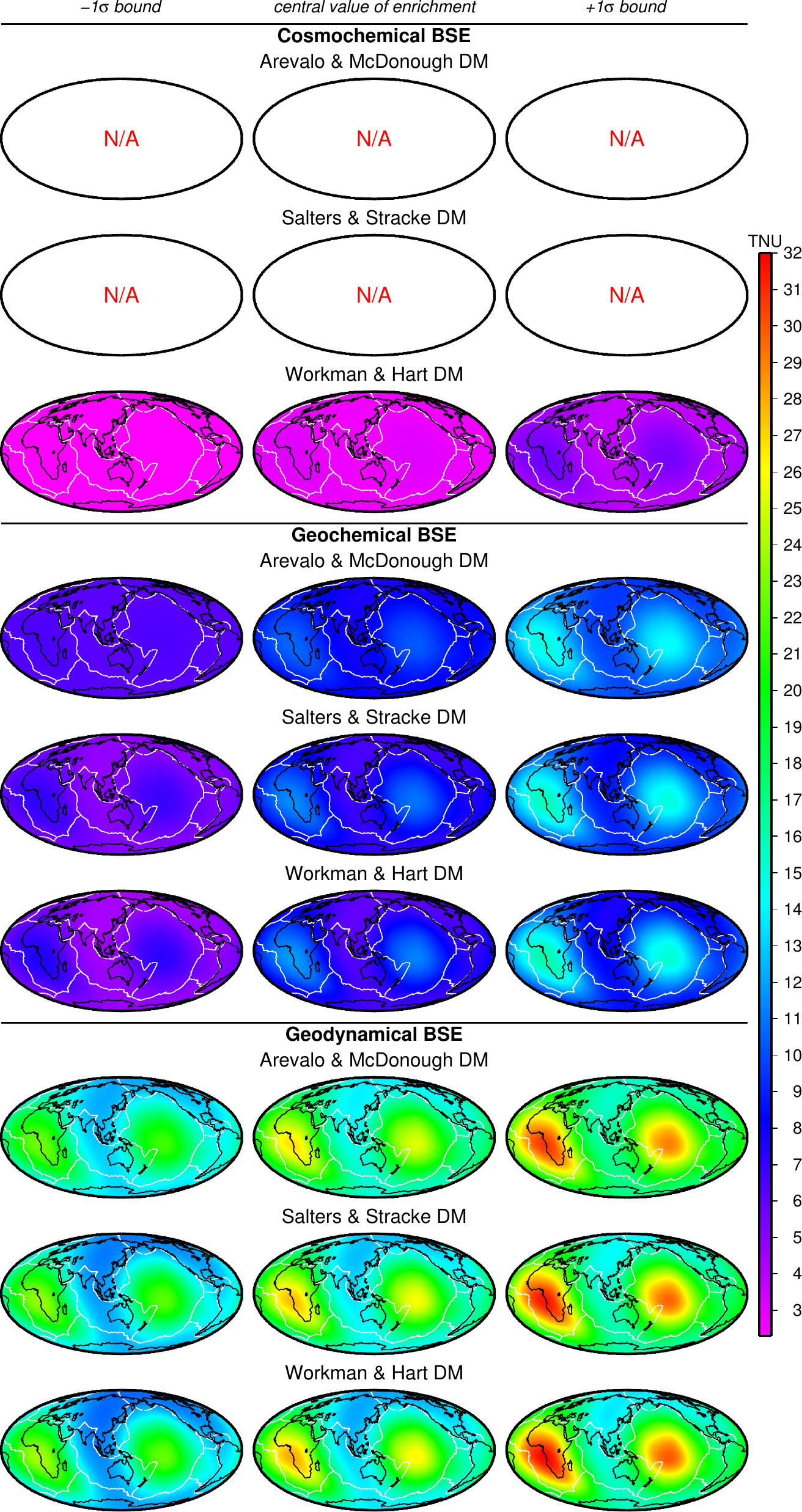}

\end{document}